\documentclass{iaus}
\usepackage{graphicx}

  \checkfont{eurm10}
  \iffontfound
    \IfFileExists{upmath.sty}
      {\typeout{^^JFound AMS Euler Roman fonts on the system,
                   using the 'upmath' package.^^J}%
       \usepackage{upmath}}
      {\typeout{^^JFound AMS Euler Roman fonts on the system, but you
                   dont seem to have the}%
       \typeout{'upmath' package installed. iaus.cls can take advantage
                 of these fonts,^^Jif you use 'upmath' package.^^J}%
      }
  \else
  \fi

  \checkfont{msam10}
  \iffontfound
    \IfFileExists{amssymb.sty}
      {\typeout{^^JFound AMS Symbol fonts on the system, using the
                'amssymb' package.^^J}%
       \usepackage{amssymb}%
         
       \let\ge=\geqslant  
      }{}
  \fi

  \IfFileExists{amsbsy.sty}
    {\typeout{^^JFound the 'amsbsy' package on the system, using it.^^J}%
     \usepackage{amsbsy}}
    {}

\newsavebox{\astrutbox}
\sbox{\astrutbox}{\rule[-5pt]{0pt}{20pt}}

\title[Outskirts of Galaxy Clusters: intense life in the suburbs]
{Action Model of Infall into the Virgo Cluster}

\author[R.B. Tully and R. Mohayaee]%
{R. Brent Tully$^{1,2}$%
\and Roya Mohayaee$^2$}

\affiliation{$^1$Institute for Astronomy, University of Hawaii, 
Honolulu, HI 96822, USA email: tully@ifa.hawaii.edu\\[\affilskip]
$^2$Observatoire de la C\^ote d'Azur, 06304 Nice, France
email: roya@obs-nice.fr}

\pubyear{2004}
\volume{195}
\pagerange{1--8}
\date{?? and in revised form ??}
\jname{Outskirts of Galaxy Clusters: intense life in the suburbs}
\editors{A. Diaferio, ed.}
\begin{document}

\maketitle

\begin{abstract}
The observed infall of galaxies into the Virgo Cluster puts strong
constraints
on the mass of the cluster.  A non-parametric fully non-linear
description of
the infall can be made with orbit reconstructions based on Numerical
Action
Methods.  The mass of the cluster is determined to be
$1.2 \times 10^{15} M_{\odot}$.  The mass-to-light ratio for the
cluster is
found to be seven times higher than the mean ratio found across the
region
within $V=3000$ km/s.
\end{abstract}

\firstsection 
\section{Introduction}

The general problem we will consider is the reconstruction of orbits
that 
galaxies might have followed to arrive at their currently observed
positions
on the sky and in redshift.  On this occasion we will focus on the
particularly
interesting circumstances associated with infall into the Virgo
Cluster.
The program involves three distinct components. 

First, we require as complete a map as possible of the angular
positions
and velocities of galaxies.  It will be assumed that there is some
correlation
between the distribution of galaxies in the catalog and the actual
distribution
of mass, which is the parameter that really interests us.  However it
is to be
anticipated from the outset that the relationship between what we see
-- 
galaxies of various types -- and the distribution of mass may be
complex
(Tully 2003, 2004).  It is also to be appreciated that large scale
tides may
be dynamically significant so one needs a map of the distribution of
galaxies
that extends well beyond the region of immediate focus.  In the
present 
instance, our interest is in the environs of the Virgo Cluster at a
distance
of 16.8 Mpc.  Our map of the distribution of galaxies
is based on a sample of 3151 galaxies within $\sim 40$ Mpc.  The
sample is 
complete
to $0.1 L^{\star}$ within 25 Mpc at high galactic latitudes.
Selection
function corrections are made as a function of distance and `fake'
sources 
are added at low latitudes to account for missing sources in the zone
of
obscuration.

The second critical component is a catalog of accurate distances to
galaxies.
Distance measures, $d$, allow a separation of observed velocities,
$V_{\rm obs}$,
and peculiar velocities, $V_{\rm pec}$, since $V_{\rm pec}=V_{\rm
  obs}-{\rm H}_0 d$.
Here, H$_0$ is the Hubble Constant which is taken to be 80 km/s/Mpc.
For
purposes of separating $V_{\rm pec}$ from $V_{\rm obs}$ it is only
required that
distances and the Hubble Constant be on the same scale; that is, with 
averaging that takes into account large scale flows: 
H$_0=<V_{\rm obs}/d>$.  If distances are, say, systematically measured
too large
then a self-consistent value of H$_0$ will be too low.  Derived values
of
$V_{\rm pec}$ are independent of the distance scale zero-point.
Another factor
with regard to distances to appreciate is that errors are a percentage
of the
distance, 10\% to 20\% depending on the methodology, with the
consequence that
errors in the derived $V_{\rm pec}$ grow linearly with distance.  At
the Virgo
Cluster a 10\% error corresponds to $\pm140$ km/s in $V_{\rm pec}$.
This 
uncertainty is tolerable for the Virgo infall problem but for more
distant
clusters the situation would be unsatisfactory.  The distance catalog
itself is a synthesis of our own observations (Pierce and Tully, in 
preparation) that exploit the luminosity--linewidth method (Tully and
Fisher
1977) and material from the literature.  That literature material
provides
distances from a variety of methodologies: luminosity--linewidth
(Mathewson et al. 1992, Lu et al. 1993, Tully and Pierce 2000), 
Cepheid (Freedman et 
al. 2001), Tip of the Giant Branch (Karachentsev et al. 2003),
Planetary
Nebula Luminosity Function (Jacoby et al. 1990), and Surface
Brightness
Fluctuation (Tonry et al. 2001).  This latter important source is
distinguished
separately in the ensuing discussion.  In total we have distances to
almost 900
galaxies from the luminosity--linewidth and other assorted methods and
292
distances from the Surface Brightness Fluctuation method.

The third ingredient is the theoretical machinery to convert
information
on the amplitude of the peculiar velocities of galaxies into a map of
the
mass distribution.  In linear theory there is a direct relation
between 
$V_{\rm pec}$ and matter density fluctuations.  The relationship is
more complex
in the vicinity of collapsed structures, such as the Virgo Cluster.
We make use of Numerical Action Methods which we have referred to
as `Least Action'.  The numerical techniques have been discussed by 
Peebles (1989,1995), Shaya et al. (1995), and Phelps (2002).  The
procedure
allows for orbit reconstructions in highly nonlinear regimes, though 
ambiguity arises if the orbits are complex, especially if dynamical
friction
or orbital energy exchange is important.  These are not serious issues
for
the Virgo infall problem.

\section{Numerical Action Methods}

The `Action' is defined as the integral along paths of the Lagrangian,
the kinetic minus the potential energy.  The equations of motion for
particle
orbits satisfy minima of the derivatives of the action with respect to
position
and momentum.  Orbits of particles that represent saddle points of the
action
and agree with boundary conditions are also {\it physically
  plausible}.

Orbits are described by the 6 elements of phase space ($\alpha_i$,
$\delta_i$,
$d_i$, $V_i^{\alpha}$, $V_i^{\delta}$, $V_i^r$) under the influence of
the
ensemble of mass elements, $m_i$, and a time frame.  In the ensuing
discussion
we will assume a flat universe with the matter density as a fraction
of the
closure density $\Omega_m=0.21$, the vacuum energy fractional density 
$\Omega_{\Lambda}=0.79$, and H$_0=80$ km/s/Mpc. This choice of
parameters sets 
the clock.
We solve a mixed boundary value
problem constrained by the 3 well-known elements of phase space today
($\alpha_i$, $\delta_i$, $V_i^r$) and the theoretically motivated
condition
that initially the 3 components of velocity were negligible; i.e,
taken to be
zero.  Then given mass assignments to the components of our galaxy
catalog,
orbits can be constructed.  

The observed luminosities of the components of the catalog, $\ell_i$, 
provide a guide for the
mass assignments.  Given $m_i=\ell_i(m_i/\ell_i)$, the uncertainties
in $m_i$
can be transformed into uncertainties in $m_i/\ell_i$.  Hence, with 
specification
of mass-to-light values we can constrain plausible orbits.  There
would be
too many degrees of freedom if all $m_i/\ell_i$ were allowed to be
different
so, to begin, we go to the other limit and assume all objects have the 
{\it same} value, $M/L$.  

Given a choice of $M/L$ within the context of a specified cosmology we
can
constrain plausible orbits, but how can discrimination be made between
alternatives?  It is here that we make use of distance measurements.
The orbits that have been defined are constrained as boundary
conditions
to have specified positions on the sky and radial velocities at the
current
epoch.  The end point distances are an output of the model.  For a 
substantial fraction of the objects we have observed distances.  These 
cases can be used to evaluate the quality of the model through a
$\chi^2$ estimator that compares differences between model and
observed
distances (the comparison is done in the logarithms using distance
moduli).
\begin{center}
\begin{figure}
\includegraphics[width=120mm]{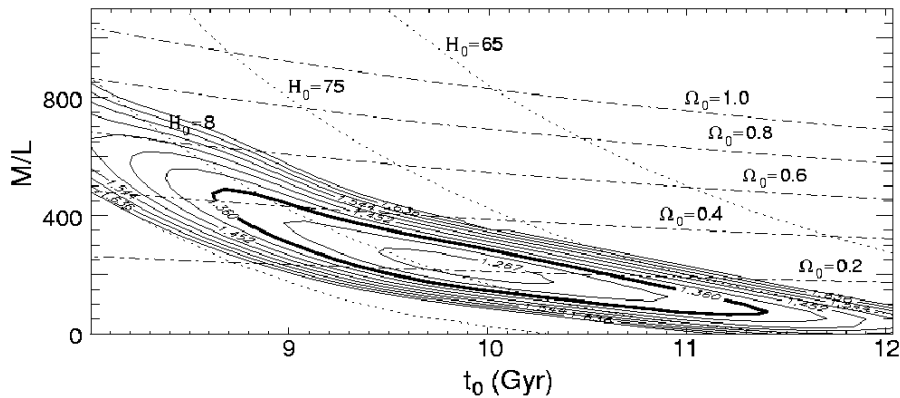}
\caption
{
Contours of constant $\chi^2$ from an Action analysis of a catalog
restricted to $V<3000$ km/s. In this case, $\Omega_{\Lambda}=0$ is 
assumed. The heavy contour represents the $2\sigma$ confidence level.
}
\label{fig1}
\end{figure}
\end{center}
Figure 1 provides an example of $\chi^2$ contours in the domain of
$M/L$ vs.
$t_0$, the age of the universe.  The analysis presented here assumed 
$\Omega_{\Lambda}=0$ and the
distance scale zero point was $10\%$ above our current preference.
These 
differences particularly affect the time domain and, at the $10\%$
level,
the $M/L$ values but do not affect the validity of the $\Omega_m$
estimates.  

It is to be emphasized what one should not and what one should take
seriously 
with the
Action models.  One should {\it not} take seriously the specific orbit
reconstructions.  The orbits that are derived are physical but not
unique.
The uniqueness problem becomes important for large-scale
reconstruction where a different method which guarantees uniqueness, 
 can be employed (Frisch et al. 2002).
One {\it should} take seriously the constraints on mass.  A thought
experiment
can make this point clear.  Suppose we consider two cases of
interaction 
between a test particle and a substantial mass where the difference
between the
two cases is simply the mass of the major attractor.  Suppose, as in
the real
world, we can measure the line-of-sight velocity of the test particle
with high
accuracy and
measure the separation between between the test particle and attractor
with
some error.  We want to estimate the mass of the attractor.  Admitting
that the
line-of-sight velocity is only a component of the true velocity we
only have
a statistical test.  However it should be clear that the measure of
the
separation between particles provides a measure of the mass.  For a
specified
infall velocity, if the attractor has more mass then the test particle
has
to be farther away.  The Action models are responding to this effect.
If there
is generally more mass, then for fixed velocities and within a
specified 
cosmological model, objects are farther apart.  Our distance
observations are
distinguishing between the mass possibilities.  

We are careful to demand that the cosmological model be specified
because,
as is clear from Fig.~1, there is a trade off between mass and time.
A model
with more mass and less time can duplicate a model with less mass and
more 
time.  Essentially the effect of 
$\Omega_{\Lambda} > 0$ is to change the clock.

\section{Virgo Infall}

Probably of all the problems we could consider with the Action
machinery,
the simplest application is in connection with infall into the Virgo
Cluster.
Here we are giving consideration to the first approach infall of
essentially
`test particles' toward a dominant mass.  Near to the cluster, the
radial
motions toward the cluster (reaching 1500 km/s) are much larger than
any
transverse motions ($\sim 100$ km/s).  Since the orbits are
essentially 
radial, on first approach, and have not yet reached the cluster there
are 
no serious concerns about orbital
energy exchange or dynamical friction.  The interactions involve only
small
perturbations on a two-body encounter.

We begin with Action model parameters that provide a reasonable
$\chi^2$
fit to distance measures throughout the region with $V < 3000$ km/s.
Initially, a single value of $M/L$ is taken for all components.
Fig.~1
provides an example of a good solution.  In the present study, we
assume a
flat cosmological model with $\Omega_m h^2 = 0.135$ (where $h={\rm
  H}_0/100$)
consistent with microwave background measurements (Spergel et
al. 2003) 
but take the slightly nonstandard values $h=0.80$ and $\Omega_m=0.21$.

As we have reported earlier (Shaya et al. 1995), good fits to the
ensemble
of the data within $V=3000$ km/s with a single $M/L$ value assigned to
all
components requires a low $M/L \sim 200 M_{\odot}/L_{\odot}$.  However
a
part of the data is poorly fit with this solution.  This model implies
a
sufficiently modest mass for the Virgo Cluster that it cannot generate
the
large infall velocities that are observed.

The so-called `triple-value' region (Tonry and Davis 1981) around a
cluster 
provides a signature of
infall that depends on the mass of the cluster.  Theoretical models 
demonstrate the properties of infall in 3-dimensions (Bertschinger
1985).  The structure
is simple in the volume within the zone of first collapse in to the
radius
of a caustic which is the outer extent of orbits of objects that have
passed
through the cluster once and reached second turnaround.  Within this
zone
between outer (first turnaround) and inner (second turnaround) radii
the
infall pattern translates to the now familiar `triple-value' wave,
whereby
there are three locations that give the same velocity.  In such
instances,
one of the locations lies outside the first turnaround -- or zero
velocity
surface for the cluster -- and the other two are within the collapse
region.
There are two infall locations that give the same velocity because of
a 
line-of-sight
geometrical effect.  Since the motions represent infall toward the
cluster,
objects on the front side of the cluster have higher velocities than
the 
cluster mean and objects on the back side of the cluster have lower
velocities
than the cluster mean.

There are two key points to understand in connection with the present
study.
The first is the obvious point that {\it the amplitude of the
  triple-value
wave depends directly on the mass of the cluster.}  Hence, if one is 
convinced that an object is within the infall region (rather than at
the
location outside the zero-velocity surface) then the mass of the
cluster
needs to be sufficient to explain the velocity of that object.  In
many
cases in which the velocities are similar to that of the cluster the
demand
will not be too great.  The challenge is to explain the extreme cases.

The second point to appreciate is that the Virgo Cluster is near
enough that
we generally can unambiguously determine whether a target is within or 
outside the infall region.  We may not be able to distinguish between
the
two infall options but that is of minor consequence.  The essential
point 
is that if the object is within the infall region then that object in
itself
puts a minimal demand on the mass interior to its position within the
infall
domain.

The general domain of the Virgo infall region is demonstrated in
Figure 2.
It is observationally determined that the outer radius of
first turnaround 
is about $28^{\circ}$ from the center of the cluster (a radius of 8
Mpc)
while the caustic of the second turnaround is at about $6^{\circ}$
(1.8 Mpc).
In Fig.~2, all the objects that are highlighted (plus others) except
for the
object 102 are within the Virgo Cluster zero-velocity surface.  It is
seen 
that many of these objects are to the left of the cluster in this
plot.
These objects lie in the `Virgo Southern Extension' which is a cloud
of 
galaxies impinging on the cluster at the present epoch.
\begin{center}
\begin{figure}
\includegraphics[width=120mm]{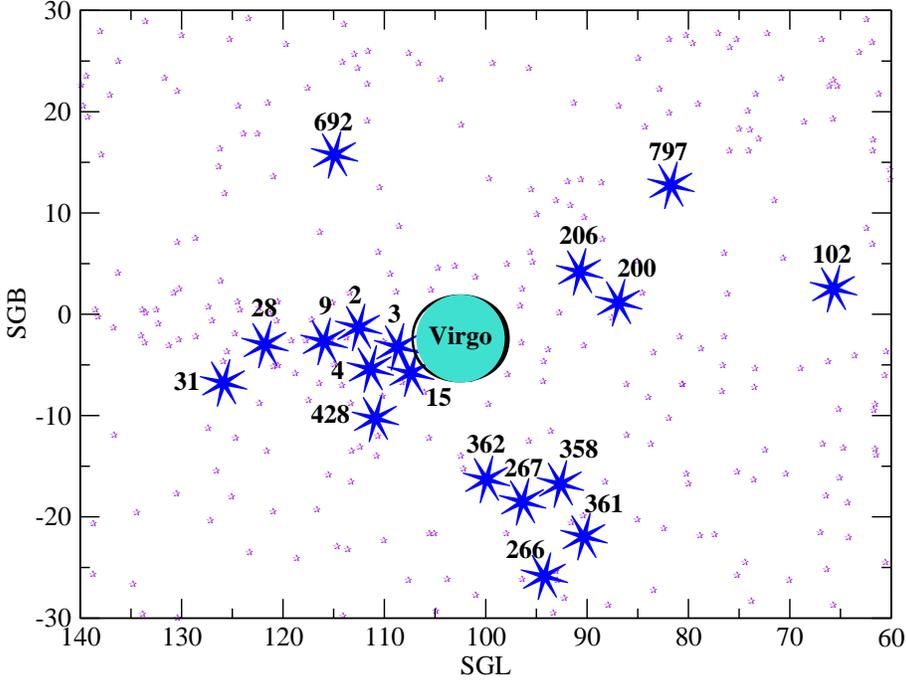}
\caption{
Projection on the sky of the Virgo Cluster and surrounding region.
The 
various symbols indicate the positions of objects projected in the
vicinity
of the cluster and those noted by large stars and numbered are
particularly 
`interesting' from the standpoint of the Virgo infall problem.  The
infall
region extends to $\sim 28^{\circ}$ around the cluster, encompassing
all but 
the object labelled 102 on this plot.
\label{fig2}}
\end{figure}
\end{center}

The objects highlighted in Fig.~2 have received
special attention because they have good distance measures and many of
them 
provide strong constraints on the cluster mass.  An example is
presented in
Figure 3.  The two panels illustrate the triple-value wave in
the
direction of a specific group of galaxies (object 4 in Fig.~2: the
11-4 Group 
in Tully 1988)
with the only difference between the two cases being the Virgo Cluster
mass
assignment.  In both cases, the Action model was solved assuming 
$M/L=125 M_{\odot}/L_{\odot}$ as a baseline.  In the case shown on the
right,
however, the mass of the clusters dominated by early type systems (for
present
purposes that means the Virgo Cluster) was augmented by 7 to 
$M/L=875 M_{\odot}/L_{\odot}$.  The mass of the cluster was increased
from
$1.6 \times 10^{14} M_{\odot}$ (left panel) to $1.2 \times 10^{15}
M_{\odot}$
(right panel).  The measured distance and velocity of the 11-4 Group is
indicated
by the point with error bars.  The group of stars indicate the
distance and
velocity of the Virgo Cluster as modelled by a series of Action
trials.
The `x' indicate distances and velocities determined for the 11-4
Group with
the Action trials.  The points that describe waves rising from zero
distance
and velocity to 40--45 Mpc at 3000 km/s are the orbital end points of
test 
particles scattered along the line-of-sight of the 11-4 Group.

\begin{figure}
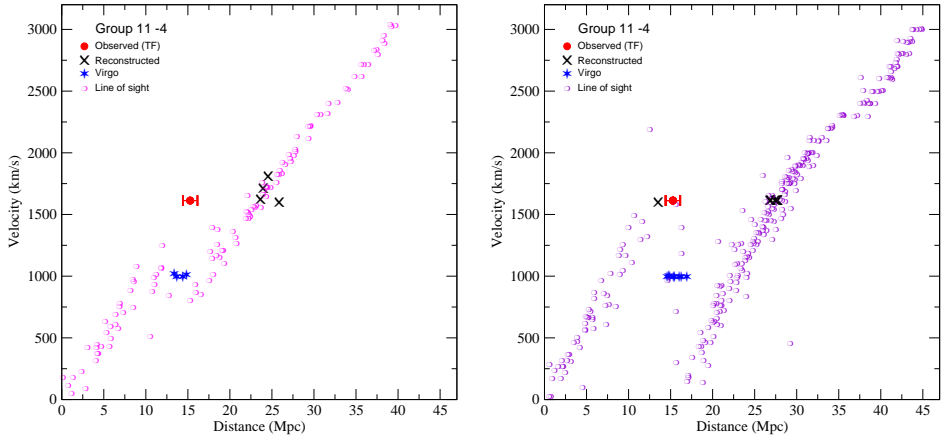

\begin{eqnarray}
\includegraphics[width=60mm]{triplevaluedA1-ls4-article.eps}\quad
&
\includegraphics[width=60mm]{triplevaluedA7-ls4-article.eps}
\nonumber
\end{eqnarray}
\caption{
The left plot is
the (limited) triple-value wave associated with the line-of-sight
directed
at the infall group 11-4 in action models with $M/L=125
M_{\odot}/L_{\odot}$
for all objects.  In this case the mass of the Virgo Cluster is 
$1.7 \times 10^{14} M_{\odot}$.  The mean Virgo Cluster distance and
velocity
given by several Action trials is indicated by the stars.  The
locations
found for the 11-4 group in different trials with the Action models
are given 
by the `x'.  The point
with error bars indicate the {\it observed} position and velocity of
the group.
The right plot is
the (much more extensive) triple-value wave along the same line of
sight 
toward the 11-4 group generated by action models still with 
$M/L=125 M_{\odot}/L_{\odot}$ for most objects but now with the mass
of the
Virgo Cluster (and other E/S0 groups in the catalog) amplified by a
factor 7.
The Virgo Cluster is now given a mass of $1.2 \times 10^{15}
M_{\odot}$.
Symbols have the same meaning as in the left panel.
\label{fig3}}
\end{figure}

There is a clear difference between the two figures in the amplitude
of the
triple-value wave in the vicinity of the Virgo Cluster. This difference
is a 
consequence of the different cluster mass assignments.  The wave
defined by the
test particles shows where {\it objects could lie} in the context of
the
specified model.  In the right panel of Fig.~3, the data point 
representing the 11-4
Group lies
on the triple-value wave, hence the model accommodates this datum.  In
the left panel,
however, the same data point lies well away from the much more modest
triple-value wave.  We are forced to conclude that either the data
point is
unreliable or the model shown in this figure is bad.  The `x' symbols
that
illustrate the positions found for the 11-4 Group in the Action models
reinforce the point.  In the left panel, the Action models put the group in
the
vicinity of 25 Mpc, in expansion away from the cluster, rather than at
it's
measured distance of $15.3 \pm 0.9$ Mpc.  In the right panel, the Action models 
frequently put the group at a similar background position but on an
occasion
a saddle point of the action was chosen in good agreement with the
observed
distance.  Clearly, in a computation of the $\chi^2$ evaluator of the 
distance agreement this latter case is strongly favored.  The
important
point to be made is that {\it in order for an infall solution to be
  available
for the 11-4 Group the mass of the cluster has to be high enough to
generate
a triple-value wave with an amplitude that catches the velocity
observed for
the group.} 

\section{Conclusions}

Similar analyses can be made of dozens of other lines-of-sight through
the
infall domain around the Virgo Cluster.  If there is a distance
measurement that
confirms that a target is inside the zero-velocity surface around the
cluster
then the line-of-sight component of its velocity puts a constraint on
the 
cluster mass.  In each case, the cluster mass must be at least enough
to
provide an Action infall solution.

We determine that a minimum but sufficient Virgo Cluster mass for the
models
is $1.2 \times 10^{15} M_{\odot}$.  This mass is $~50\%$ larger than
the
mass determined from application of the Virial theorem (Tully and
Shaya 1984).  
However, 
the Virial radius for the cluster is $\sim 0.8$ Mpc while the mass
measured
by infall is on a scale $\ge 1.8$ Mpc.  The infall analysis, which is
based
on very simple physics, gives an estimate 
of the {\it global} mass of the cluster.  It can be anticipated that
this mass
will exceed the masses determined over more limited scales by
gravitational
lensing, X-ray, or Virial studies.

Unfortunately, the analysis is not easily duplicated on other
clusters, at
least not so cleanly.  The Virgo Cluster is the only environment {\it
  near}
enough that distance measurements distinguish between infall and
expansion
regimes, and {\it massive} enough to have an extensive, well populated
infall
domain.  It provides a single good case.  What we find, though, is
that
whereas overall the Action models find preference for 
$M/L \sim 125 M_{\odot}/L_{\odot}$ and $\Omega_m \sim 0.2$, in the
Virgo 
Cluster we need $M/L \sim 900 M_{\odot}/L_{\odot}$.

\begin{acknowledgments}
This research is being carried out in close collaboration with Jim
Peebles,
Steven Phelps, and Ed Shaya.  Support for RBT is provided by JPL
Contract 
1243647. RM is supported by the European Union Marie Curie Fellowship
HPMF-CT-2002-01532.
\end{acknowledgments}

\end{document}